Title: Vibronic solitons in a (nearly-) degenerate electronic system along a 1-D chain
Authors: Mladen Georgiev (Institute of Solid State Physics, Bulgarian Academy of
    Sciences, 72 Tsagigradsko Chaussee, 1784 Sofia, Bulgaria)
Comments: 7 pages and 1 figure pdf format
Subj-class: cond-mat

We consider the formation and stability of the vibronic polaron arising in a degenerate or nearly degenerate electronic system coupled to an appropriate vibrational mode. We define the electron-phonon coupling as a mixing of the electronic states by the phonon mode. Our investigation is similar to that carried out recently by Dennis Clougherty though by introducing a different coupling operator. This extends the vibronic coupling range beyond the limits set by the difference in electronic density of the progenitor bands thereby widening the applicability range of the coupled soliton.

1. Rationale

The problem of a vibronic polaron arising through Jahn-Teller (JT) or Pseudo-Jahn-Teller (PJT) coupling in a degenerate or nearly-degenerate two-band electronic system has attracted considereble attention for some time due to its general theoretical significance on the one hand and to its presumed applicability to a number of practically important systems on the other [1]. Theoretically these species can be either itinerant or bound (localized). Due to its math complexity, the itinerant problem has only been dealt with by applying the Variational Ansatz to the Hamiltonian and the field operator in second quantization terms [2,3]. Both works define a coupling term linear in the phonon coordinate whereat the factor is purely electronic in character. The former paper defines the factor as difference in electronic density between the two progenitor electronic bands,[2] the latter one defines it by pair products of electron ladder operators signifying annihilation of an electron in band "one" and subsequent creation of another electron in band "two" [3]. While the density-based (band-diagonal) definition has been criticized for an aledged non-genuine mixing character,[4] it has gained popularity because of an early variational work on the itinerant JT polaron [2]. Unlike it, the latter type (band off-diagonal) has widely been used for dealing with JT or PJT problems in chemistry [5-7].

Recently, Clougherty and Foell considered a 1-D model of a particle in a doubly (or nearly) degenerate electronic band coupled to an elastic distortion [8]. They showed that the equations of motion of the particle reduce to coupled nonlinear Schrödinger equations. For an interband electron-phonon coupling stemming from local JT or PJT interactions, multicomponent self-trapped polaron solutions – vectror polarons – are described and classified and their phase diagram documented.

## 2. Model

Following [8] we define a model Langrangian of an electron embedded in a dimodal deformable medium:

$$L = L_e + L_{ph} + L_{ep} \quad (1)$$

with

$$L_e = \tfrac{1}{2} i h (\Psi^\dagger \partial_t \Psi - \Psi^T \partial_t \Psi) - (h^2/2m) \partial_x \Psi^\dagger \partial_x \Psi - \Delta \Psi^\dagger \sigma_z \Psi \quad (2)$$

$$L_{ph} = - \sum_i \tfrac{1}{2} \kappa_i \phi_i^2 \quad (3)$$

$$L_{ep} = g_0 \phi_0 \Psi^\dagger \Psi + g_1 \phi_1 \Psi^\dagger \sigma_x \Psi \quad (4)$$

where $\Psi$ ($\Psi^T = (\psi_1 \; \psi_2)$) is the spinor field operator, $\Delta$ is the energy gap between $\psi_2$ and $\psi_1$, finite for nearly-degenerate and vanishing for degenerate electronic states. $\phi_i$ is the coupled lattice distortion, $\kappa_i$ is the stiffness, $g_0$ is the electron-phonon coupling constant (Holstein mode), $g_1$ is the interband mixing constant (vibronic mode), $\sigma_z = \text{diag}\{1, -1\}$, $\sigma_y = \text{off-diag}\{-i, i\}$, $\sigma_x = \text{off-diag}\{1,1\}$, $\sigma_0 \equiv 1 = \text{diag}\{1,1\}$ are the Pauli matrices [6]. Clougherty and Foell have chosen $\sigma_z$ to construct $L_{ep}$ [8] we choose $\sigma_x$ instead. Our choice is based on the premise that while $\sigma_z$ generates diagonal rearrangements only, $\sigma_x$ produces a genuine off-diagonal (interband) mixing of the electronic operators. While the $\sigma_z$ case has been explored in detail by Clougherty and Foell, we aim at the $\sigma_x$ case as the goal of our present work (See reference [8] for greater details. To facilitate the comparison, we preserve the original notations of the latter reference wherever possible.)

For the sake of completeness we remind that while the Holstein mode preserves the site symmetry, the vibronic mode breaks the local symmetry. Next we assume a classical lattice in the adiabatic approximation to perform minimization in $\phi_0$ and $\phi_1$ giving

$$\phi_0 = (g_0 / \kappa_0)(\Psi^\dagger \Psi) = (g_0/k_0)(\psi_1^* \psi_1 + \psi_2^* \psi_2) \quad (5)$$

$$\phi_1 = (g_1/k_1)(\Psi^\dagger \sigma_x \Psi) = (g_1/k_1)(\psi_1^* \psi_2 + \psi_2^* \psi_1) \quad (6)$$

In the adiabatic approximation, the lattice kinetic energy is discarded. There are a number of adiabatic investigations of linear atomic chains [9]. Eliminating the distortions from the relevant equations of motion, we arrive at the following nonlinear Schrödinger equations:

$$i h \partial_t \psi_1 - \Delta \psi_1 + \tfrac{1}{2}(h^2/m)\partial_x^2 \psi_1 + [\gamma_0^4(|\psi_1|^2 + |\psi_2|^2) + \gamma_1^4(\psi_2^* \psi_1 + \psi_1^* \psi_2)]\psi_1 = 0 \quad (7)$$

$$i h \partial_t \psi_2 + \Delta \psi_2 + \tfrac{1}{2}(h^2/m)\partial_x^2 \psi_2 + [\gamma_0^4(|\psi_1|^2 + |\psi_2|^2) + \gamma_1^4(\psi_2^* \psi_1 + \psi_1^* \psi_2)]\psi_2 = 0 \quad (8)$$

where

$$\tfrac{1}{2} \gamma_0^4 = g_0^2 / 2\kappa_0 = \varepsilon_{JT0} \tag{9}$$

$$\tfrac{1}{2} \gamma_1^4 = g_1^2 / 2\kappa_1 = \varepsilon_{JT1} \tag{10}$$

are the JT energies of the Holstein mode (self-modulation) and the vibronic mode (cross-modulation), respectively [7].

We next seek Riemann solutions to the nonlinear equations in the form

$$\psi_j = \sqrt{[\,|E - \Delta|\,/\,\nu\,]} \exp\{\,i\,[\,\upsilon x - (E - \tfrac{1}{2}\,m\upsilon^2)\,t\,]\,\}\,r_j\,(\sqrt{[2|E - \Delta|]}(x - \upsilon t)) \tag{11}$$

where for brevity

$$\nu = \gamma_0^4 = g_0^2 / \kappa_0 \tag{12}$$

$$\eta = \gamma_1^4 = g_1^2 / \kappa_1 \tag{13}$$

Inserting into (7) and (8) a set of coupled nonlinear equations for the envelope functions $r_j$ results:

$$r_1(\xi)'' + (E - \Delta)/|E - \Delta| \times r_1(\xi) + r_1(\xi)\bigl(r_1(\xi)^2 + r_2(\xi)^2 + 2(\eta/\nu)\,r_2(\xi)\,r_1(\xi)\bigr) = 0 \tag{14}$$

$$r_2(\xi)'' + (E + \Delta)/|E - \Delta| \times r_2(\xi) + r_2(\xi)\bigl(r_2(\xi)^2 + r_1(\xi)^2 + 2(\eta/\nu)\,r_1(\xi)\,r_2(\xi)\bigr) = 0 \tag{15}$$

wherein $\xi = \sqrt{[2|E - \Delta|]}\,(x - \upsilon t)$. Self-trapped solutions exist for $E < 0$ and $|E| > \Delta$ only. Setting $\omega^2 = (|E| - \Delta)/(|E| + \Delta)$ and $\beta = \eta/\nu$ we transform the latter equations into

$$r_1'' - r_1 + r_1(r_1^2 + r_2^2 + 2\beta r_2 r_1) = 0 \tag{16}$$

$$r_2'' - \omega^2 r_2 + r_2(r_2^2 + r_1^2 + 2\beta r_1 r_2) = 0 \tag{17}$$

with $0 \leq \beta \leq 1$, the condition for a stable polaron solution. Our subsequent analysis goes along the Reference [8] lines. We note that the use of the interband Pauli matrix in (4) does not apparently result in any essential change of course of the qualitative analysis for the nonlinear equations (14) through (17). As an exception though, we stress the different meanings of our self- and cross- modulation coefficients in (12) and (13).

*Polaron types.* We distinguish between three polaron types:

Type I. Exhibits identical envelope amplitudes $r_1(\xi) = r_2(\xi)$ and an orbital mixing that is constant over the spatial extent of the polaron cloud. It occurs when $\omega = 1$ and $0 \leq \beta < 1$. We note that the former equality implies $\Delta = 0$, that is, a JT situation where the uncoupled electron states are degenerate. The latter inequality implies that the interband mixing strength is inferior to Holstein's.

Type II. Exhibits asymmetric envelope amplitudes $r_1(\xi) \neq r_2(\xi)$ which occurs when $\omega \neq 1$, a PJT situation. In the wave-daughter wave approximation (e.g. $|r_1(\xi)| \gg |r_2(\xi)|$) to leading order of the daughter wave amplitude, $\omega = \frac{1}{2}[\sqrt{(1+8\beta)} - 1]$. The polaron is unbound unless $\gamma_1^2 + \gamma_0^2 > \sqrt{(24\Delta)}$. Not surprisingly there is a critical coupling strength for the distortion to develop as it is for the molecular PJT effect [6,7]. It also follows from Reference [8] that type II arises only when $\beta \geq (1+\sqrt{3})/6$.

Type III. Results for $\beta = 1$ (same JT energies $\varepsilon_{JT0} = \varepsilon_{JT1}$ and therefrom same distortion strengths $\gamma_1 = \gamma_0$) and $\omega = 1$ (JT degeneracy). For obtaining an envelope solution we set $\omega = \beta = 1$ and then sum up (16) and (17) to get:

$$(r_1 + r_2)'' - (r_1 + r_2) + (r_1 + r_2)^3 = 0 \tag{18}$$

which is readily solved. Indeed, rewriting (18) by setting $y = r_1 + r_2$, then multiplying both sides by $y'$ and integrating, we get:

$$y' = \pm 2\sqrt{(-\tfrac{1}{2} y^2 + \tfrac{1}{4} y^4 + C)} \tag{19}$$

(+ for the ascending branch, − for the descending branch of the $y(\xi)$ graphics.) Depending on the value of the first integration constant C we obtain different classes of solutions. In as much as the function under square root should be nonnegative, we have $-\tfrac{1}{2} y_0^2 + \tfrac{1}{4} y_0^4 + C \geq 0$. We remind that $\xi = \sqrt{[2|E-\Delta|]}\,(x - \upsilon t)$.

Just to see where we are, we first set C = 0. This implies $y_0 \geq \sqrt{2}$. Integrating we arrive at:

$$\xi - \xi_0 = \pm \int_{y_0}^{y} dy / 2\sqrt{(-\tfrac{1}{2} y^2 + \tfrac{1}{4} y^4)} \tag{20}$$

leading to

$$\mu \sin[\sqrt{2}\,(\xi - \xi_0)] = \sin\{\sin^{-1}(\sqrt{2}/y) - \sin^{-1}(\sqrt{2}/y_0)\} \tag{21}$$

After certain manipulations we obtain (C = 0):

$$y = \sqrt{2} / \{\mu\sin[\sqrt{2}\,(\xi-\xi_0)]\sqrt{(1-2/y_0^2)} \pm \tfrac{1}{2}\sqrt{\{[\sin[\sqrt{2}\,(\xi-\xi_0)]\sqrt{(1-2/y_0^2)}]^2}$$

$$- \sin^2[\sqrt{2}\,(\xi-\xi_0)] - (2/y_0^2)\}} \tag{22}$$

Inserting back in (16) or (17) we separate the partial envelope amplitudes $r_1$ and $r_2$. However, it is apparent from equations (16) and (17) that $r_1 = r_2$. The resulting solutions are shown in Figure 1.

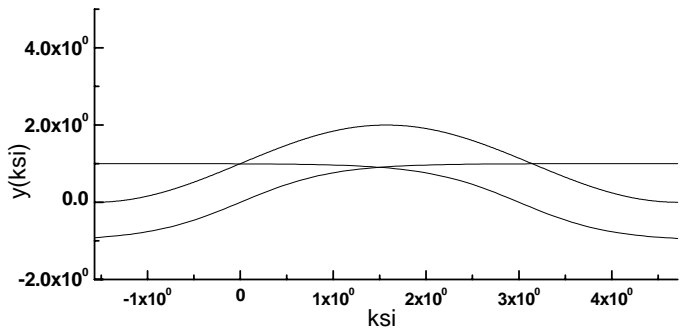

Figure 1. Graphics illustrating two solutions for the envelope function of type III vibronic solitons obtained at C = 0 (upper) and C = ¼ (lower), respectively. The upper one is formed by part of the periodic elliptic sine function at k = 0, the lower one is composed of two opposite branches of the aperiodic solution at k = 1. There is a continuum of waveforms generated by the elliptic functions in-between at 0 < k < 1.

For $C \neq 0$ we set $C = \frac{1}{2} y_0^2 - \frac{1}{4} y_0^4 = \frac{1}{2} y_0 (1 - \frac{1}{2} y_0)$ to secure that the derivative is vanishing at $\xi_0$, the expected extremal point of the soliton cloud. Now the first derivative is rearranged to read:

$$y' = \pm 2\sqrt{(-\frac{1}{2} y^2 + \frac{1}{4} y^4 + C)}$$

$$= \pm 2\sqrt{[\{y^2 - [1 + \sqrt{(1 - 4C)}]\}\{y^2 - [1 - \sqrt{(1 - 4C)}]\}]} \tag{23}$$

meaningful at $C \geq \frac{1}{4}$. The obtained derivative $y'(\xi)$ will be integrated in well tabulated elliptic integrals. Indeed, integrating $y'$ we get

$$\xi - \xi_0 = \pm \{1 / \sqrt{[1 + \sqrt{(1 - 4C)}]}\} \int_{y_{10}}^{y_1} dy_1 / \sqrt{[(1 - y_1^2)(1 - k^2 y_1^2)]} \tag{24}$$

where

$$y_1^2 = y^2 / [1 - \sqrt{(1 - 4C)}], \quad k^2 y_1^2 = y^2 / [1 + \sqrt{(1 - 4C)}],$$

$$k^2 = [1 - \sqrt{(1 - 4C)}] / [1 + \sqrt{(1 - 4C)}], \tag{25}$$

$$\sqrt{(1 - 4C)} = (1 - k^2) / (1 + k^2), \quad 1 / \sqrt{[1 + \sqrt{(1 - 4C)}]} = 1 / \sqrt{[2 / (1 + k^2)]}.$$

Inserting all we arrive at:

$$\xi - \xi_0 = \pm \{1 / \sqrt{[2 / (1 + k^2)]}\} \int_{y_{10}}^{y_1} dy_1 / \sqrt{[\{1 - y_1^2)(1 - k^2 y_1^2)]}$$

$$\xi - \xi_0 = \pm \{1 / \sqrt{[2 / (1 + k^2)]}\}\{F(y_1, k) - F(y_{10}, k)\}, \quad (y_1 = \sin\varphi) \tag{26}$$

$$F(y, k) = \int_0^{\sin\varphi} dy / [(1 - y^2)(1 - k^2 y^2)] \tag{27}$$

is the incomplete elliptic integral of first kind [10-12]. The various classes of solutions being dependent on the integration constant C, they are now controlled by the alternative parameter $\varphi$. We see that the C = 0 solution is obtained from equations (26) at k = 0 (cf. eq. (20)):

$$\sqrt{2}\,(\xi - \xi_0) = \pm \sin^{-1} y_1 - \sin^{-1} y_{10}$$

while another noticeable $C = \frac{1}{4}$ solution obtains at $k = 1$:

$$\xi - \xi_0 = \pm \tanh^{-1} y_1 - \tanh^{-1} y_{10}$$

At $C \neq 0, \frac{1}{4}$ the envelope function is expressed by means of Jacobi's elliptic sine snu:[12]

$$\pm \sqrt{[2/(1+k^2)]}\,(\xi - \xi_0) = u \equiv {}_{\sin\varphi_0}\!\int^{\sin\varphi} dy_1 / \sqrt{[(1-y_1^2)(1-k^2 y_1^2)]} \qquad (28)$$

where snu = $\sin\varphi$. Graphics of type III polarons are shown in Figure 1.

*Phase diagram.* The $\gamma_0$–$\gamma_1$ phase diagram of the three types of polaron solution can also be deduced as in Reference [8]. The abscissa $\gamma_0$ can be regarded as Holstein's mode axis, the ordinate $\gamma_1$ is the interband mode axis. $\gamma_1 = \gamma_0$ is the border line below which Holstein's coupling predominates, above it interband mixing predominates. In effect, $\gamma_1 = \gamma_0$ is the locus of points for type III polarons. The border between ranges I and II is $\gamma_1^2 = \sqrt{(\gamma_0^4 + 24\Delta)} - \gamma_0^2$. There are no stable bound polarons for $\gamma_1 > \gamma_0$, type I are stable throughout at $\gamma_1 \leq \gamma_0$, but type II have a lower binding energy within a wedge-shaped region. Type I polarons continue onto the locus of points $\gamma_1 = \gamma_0$ to become type III polarons.

### 3. Concluding remarks

Earlier work on vibronic polarons has concentrated on the adiabatic approximation to bound polarons mainly [9]. There have also been two studies using the Variational Ansatz to deal with the itinerant species as well [2,3]. The present work is the second in a series of at least two papers stimulated by the same source [8]. A solitonic approach is applied to the shape of the distortion cloud surrounding a moving electron in a deformable medium. The assumed model being one of a deformable dimodal 1-D atomic chain, it can be extended to higher dimensions or modality, but it can also serve as the theoretical background to understanding quasi-1-D structures in real materials [13].

There are two alternative approaches to the vibronic soliton depending on the form of the coupling to the symmetry-breaking vibrational mode, diagonal and off-diagonal with respect to the electronic band operators, respectively. They lead to complementing physical realities of the resulting species. While the former has been explored in reference [8], the latter is studied presently. An important consequence of our work seems to be that both approaches lead to convergent results. This conclusion has to be checked against the background of studies by the variational Ansatz [3].

Indeed, what we see on solving the nonlinear equations for the envelope functions $r_j(\xi)$ is a bell-shaped wave packet which moves along the chain. As an example, we point to the relevant Figure 1 for type III polarons. The resulting waveshapes are to be compared with earlier numerical calculations of coupled phonon amplitudes in momentum space [3].

Before closing, it is worth mentioning that similar soliton waveforms will be obtained if both coupling strengths (Holstein's & JT) are vanishing in a degenerate electron system. Indeed, in this case $\omega = 0$ and $\beta = \eta / \nu \to 1$ so that the analysis leading to eqs. (18) still holds true. We conclude that the type III polaron waveform is weakly dependent on the vibrational coupling of comparable magnitudes in case of an electronic degeneracy. Under these conditions the vibronic soliton waveform may be regarded as an inherent electronic property.

We believe our present results may be found useful for defining the realm of vibronic polaron physics.

*Acknowledgement*. We are indebted to Dr. Dennis Clougherty for mailing available an electronic copy of his interesting paper which stimulated largely our present work.

References

[1]  J.B. Goodenough, Annual Rev. Mater. Sci. **28**, 1-27 (1998).
[2] K.-H. Höck, H. Nickisch and H. Thomas, Helvetica Phys. Acta **56**, 237-243 ( 1983).
[3] M. Georgiev, cond-mat/0601676; D.W. Brown, K. Lindenberg, M. Georgiev, cond-mat/0602052; M. Georgiev, A.G. Andreev, M.D. Ivanovich, cond-mat/0602147.
[4] G.A. Gehring and K.A. Gehring, Rep. Progr. Phys. **38**, 1-89 (1975).
[5] I.B. Bersuker,.*Electronic structure and properties of transition metal compounds* (Wiley, New York, 1996).
[6] I.B. Bersuker and V.Z. Polinger,: *Vibronic interactions in molecules and crystals* (Springer, Berlin, 1989).
[7] I.B Bersuker, *The Jahn-Teller effect and vibronic interactions in modern chemistry* (Academic, New York, 1986).
[8] D.P. Clougherty and C.A. Foell, Phys. Rev. B **70**, 052301 (2004).
[9] For a recent submission see M. Georgiev and M. Borissov, cond-mat/0601423.
[10] M. Abramowitz, I.A. Stegun, eds. *Handbook of Mathematical Functions with Formulas, Graphs and Mathematical Tables*, NBS Math. Series, 1964. Russian translation: (Nauka, Moscow, 1979).
[11] E.Janke, F. Emde, F. Loesch, *Tafeln Hoeherer Funktionen*  (Teubner, Stuttgart, 1960). Russian translation: (Nauka, Moscow, 1964).
[12] I.S. Gradsteyn and I.M. Ryzhik, *Tablitsy Integralov,  summ, ryadov i proizvedenii*. (GIFML, Moskva, 1963). English: I.S. Gradshteyn and I.M. Ryzhik, *Table of Integrals, Series, and Products*, Corrected and Enlarged Edition (Academic Press, Orlando, 1965).
[13] A.G. Andreev, S.G. Tsintsarska, M.D. Ivanovich, I. Polyanski, M. Georgiev, A.D. Gochev, Central Eur. J. Phys. **2**, 89-116 (2004).